\documentclass[fleqn,10pt]{wlscirep}
\usepackage[utf8]{inputenc}
\usepackage[T1]{fontenc}
\usepackage{lineno}
\linenumbers

\usepackage{multirow}

\usepackage{graphicx}
\usepackage[lofdepth,lotdepth]{subfig}
\usepackage[colorinlistoftodos]{todonotes}
\usepackage[colorlinks=true, allcolors=blue]{hyperref}
\usepackage{xcolor}
\usepackage{comment}
\usepackage{soul}

\usepackage[nice]{nicefrac}

\usepackage[normalem]{ulem}

\graphicspath{{./}{figures/}}

\usepackage{scalerel}
\usepackage{tikz}
\usetikzlibrary{svg.path}

\definecolor{orcidlogocol}{HTML}{A6CE39}
\tikzset{
  orcidlogo/.pic={
    \fill[orcidlogocol] svg{M256,128c0,70.7-57.3,128-128,128C57.3,256,0,198.7,0,128C0,57.3,57.3,0,128,0C198.7,0,256,57.3,256,128z};
    \fill[white] svg{M86.3,186.2H70.9V79.1h15.4v48.4V186.2z}
                 svg{M108.9,79.1h41.6c39.6,0,57,28.3,57,53.6c0,27.5-21.5,53.6-56.8,53.6h-41.8V79.1z M124.3,172.4h24.5c34.9,0,42.9-26.5,42.9-39.7c0-21.5-13.7-39.7-43.7-39.7h-23.7V172.4z}
                 svg{M88.7,56.8c0,5.5-4.5,10.1-10.1,10.1c-5.6,0-10.1-4.6-10.1-10.1c0-5.6,4.5-10.1,10.1-10.1C84.2,46.7,88.7,51.3,88.7,56.8z};
  }
}

\newcommand\orcidicon[1]{\href{https://orcid.org/#1}{\mbox{\scalerel*{
\begin{tikzpicture}[yscale=-1,transform shape]
\pic{orcidlogo};
\end{tikzpicture}
}{|}}}}

\title{
A Machine-Learning-Ready Dataset Prepared from the Solar and Heliospheric Observatory Mission
}

\author[1,*]{Carl Shneider}
\author[1]{Andong Hu}
\author[1]{Ajay K. Tiwari}
\author[2]{Monica G. Bobra}
\author[5]{Karl Battams}
\author[1]{Jannis Teunissen}
\author[3,4]{Enrico Camporeale}

\affil[1]{Center for Mathematics and Computer Science, Multiscale Dynamics, Amsterdam, 1098 XG, the Netherlands}
\affil[2]{W.W. Hansen Experimental Physics Laboratory, Stanford University, Stanford, CA, 94305, USA}
\affil[3]{CIRES, University of Colorado, Boulder, CO, 80309, USA}
\affil[4]{NOAA, Space Weather Prediction Center, Boulder, CO, 80305, USA}
\affil[5]{US Naval Research Laboratory, Washington DC, USA}

\affil[*]{corresponding author(s): Carl Shneider (shneider.carl@gmail.com)}

\begin{abstract}

We present a Python tool to generate a standard dataset from solar images that allows for user-defined selection criteria and a range of pre-processing steps. 
Our Python tool works with all image products from both the Solar and Heliospheric Observatory (SoHO) and Solar Dynamics Observatory (SDO) missions. 
We discuss a dataset produced from the SoHO mission’s multi-spectral images which is free of missing or corrupt data as well as planetary transits in coronagraph images, and is temporally synced making it ready for input to a machine learning system. 
Machine-learning-ready images are a valuable resource for the community because they can be used, for example, for forecasting space weather parameters. 
We illustrate the use of this data with a 3-5 day-ahead forecast of the north-south component of the interplanetary magnetic field (IMF) observed at Lagrange point one (L1).
For this use case, we apply a deep convolutional neural network (CNN) to a subset of the full SoHO dataset and compare with baseline results from a Gaussian Naive Bayes classifier.

\end{abstract}

\begin{document}

\flushbottom
\maketitle

\thispagestyle{empty}


\section{Background \& Summary} \label{sec:intro}

Studies based on physics models have shown that solar magnetic field captured with magnetograms 
contain crucial information for estimating the speed of the solar wind, while the dynamical features of CMEs (angular width, initial speed, etc.) are routinely inferred from coronagraph images\cite{solwindmagneto,CMEpropsCoronagraphs}. 
Hence, it is expected that applying machine learning (ML) techniques to the high temporal coverage data of both the Solar and Heliospheric Observatory\cite{SoHO_overview} (SoHO) mission and the Solar Dynamics Observatory\cite{SDO} (SDO) mission is a feasible venture, that can potentially improve the space weather forecasting capability of current models\cite{singer01,schrijver15,schwenn06}.
With the quality of input data remaining paramount to the success of these ML-methods\cite{camporealeML,SWMF21,tenways} and to ensure reproducible scientific research\cite{snakes_spaceship, HelioML2020}, the preparation of a community-wide standard dataset with a standard software is crucial. 

At present, SoHO has provided more temporal coverage of the Sun than its successor, NASA's SDO, and has also fully covered Solar Cycle $23$ and $24$ with a suite of on-board instruments\cite{SoHO_overview} including those specific to solar imaging: the Michelson Doppler Imager\cite{SoHO_mdi} (MDI) for the solar photosphere, the Extreme ultraviolet Imaging Telescope\cite{SoHO_eit} (EIT) for the stellar atmosphere to low corona, and the Large Angle and Spectrometric Coronagraph\cite{SoHO_lasco} (LASCO) covering the corona from $1.5-30~R_s$, detailed in Table~\ref{tab:SoHO}. 
Recently, a white-light coronal brightness index (CBI), constructed from the full LASCO~C2 mission archive, was used to explore correlations between the solar corona and several geophysical indices\cite{battams20}.
Although SDO has even higher resolution and cadence, SoHO continues to uniquely provide coronagraph products and serves as a mission critical backup to the SDO for solar flare and CME forecasting in the event of SDO failure.

\begin{figure}[h!] 
\centering
\subfloat[Subfigure 1 list of figures text][MDI]{
\includegraphics[width=0.35\textwidth,scale=0.6]{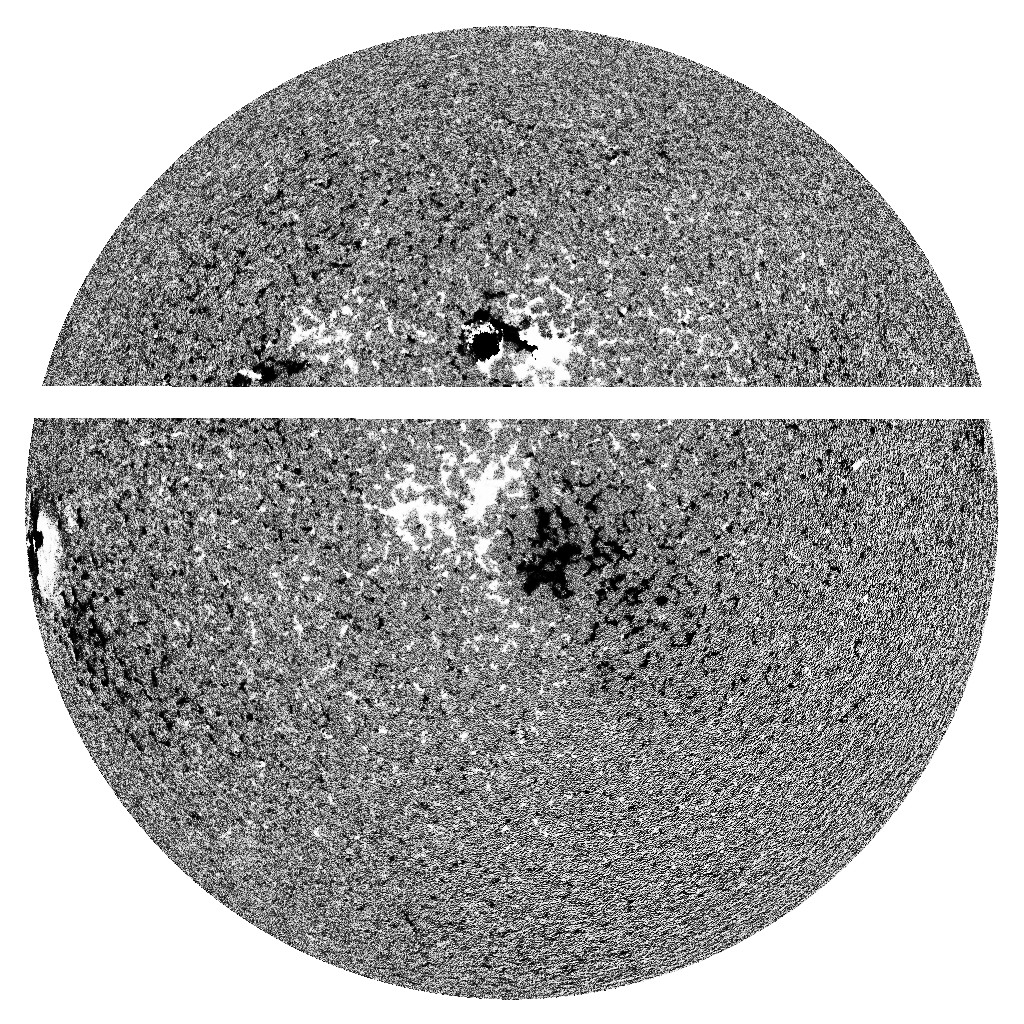} 
\label{fig:MDI}}
\subfloat[Subfigure 2 list of figures text][EIT 195]{
\includegraphics[width=0.35\textwidth,scale=0.6]{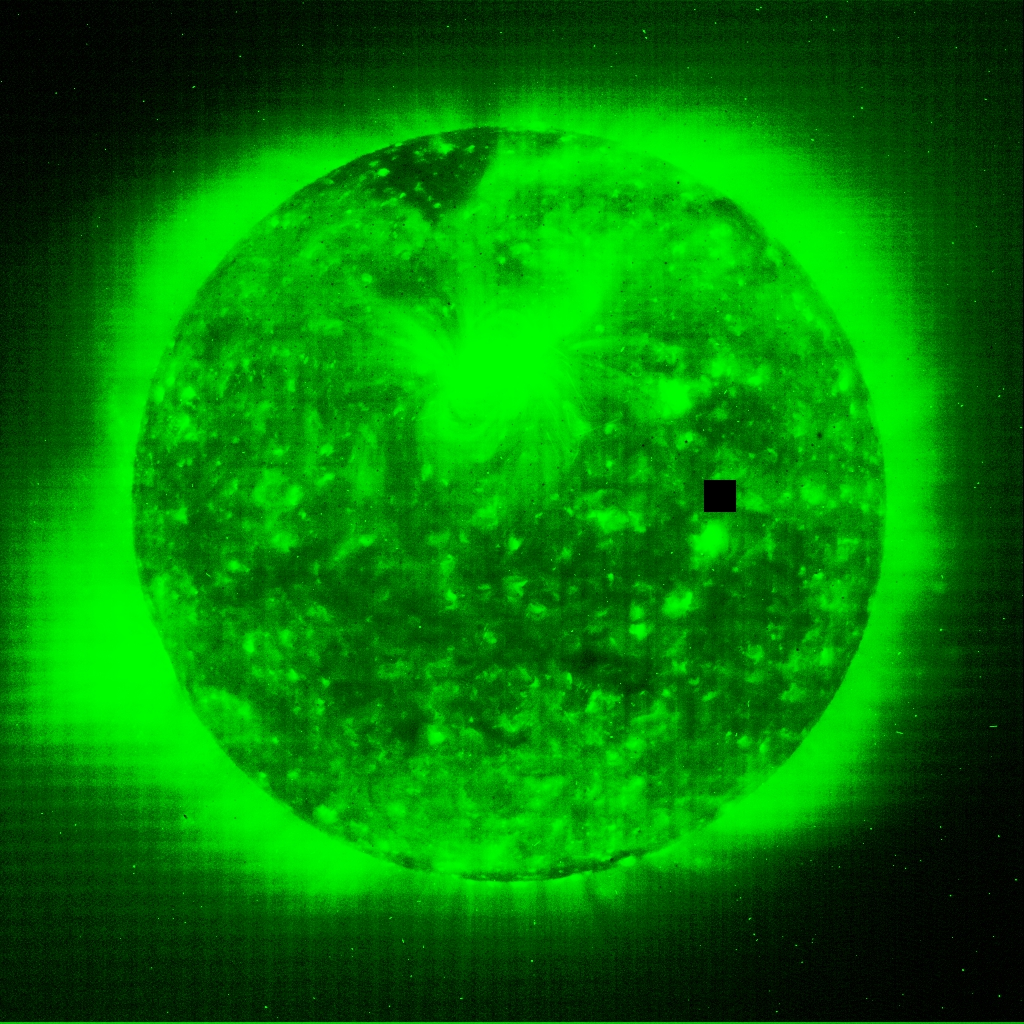}
\label{fig:EIT195}}
\qquad
\subfloat[Subfigure 3 list of figures text][LASCO C2]{
\includegraphics[width=0.35\textwidth,scale=0.6]{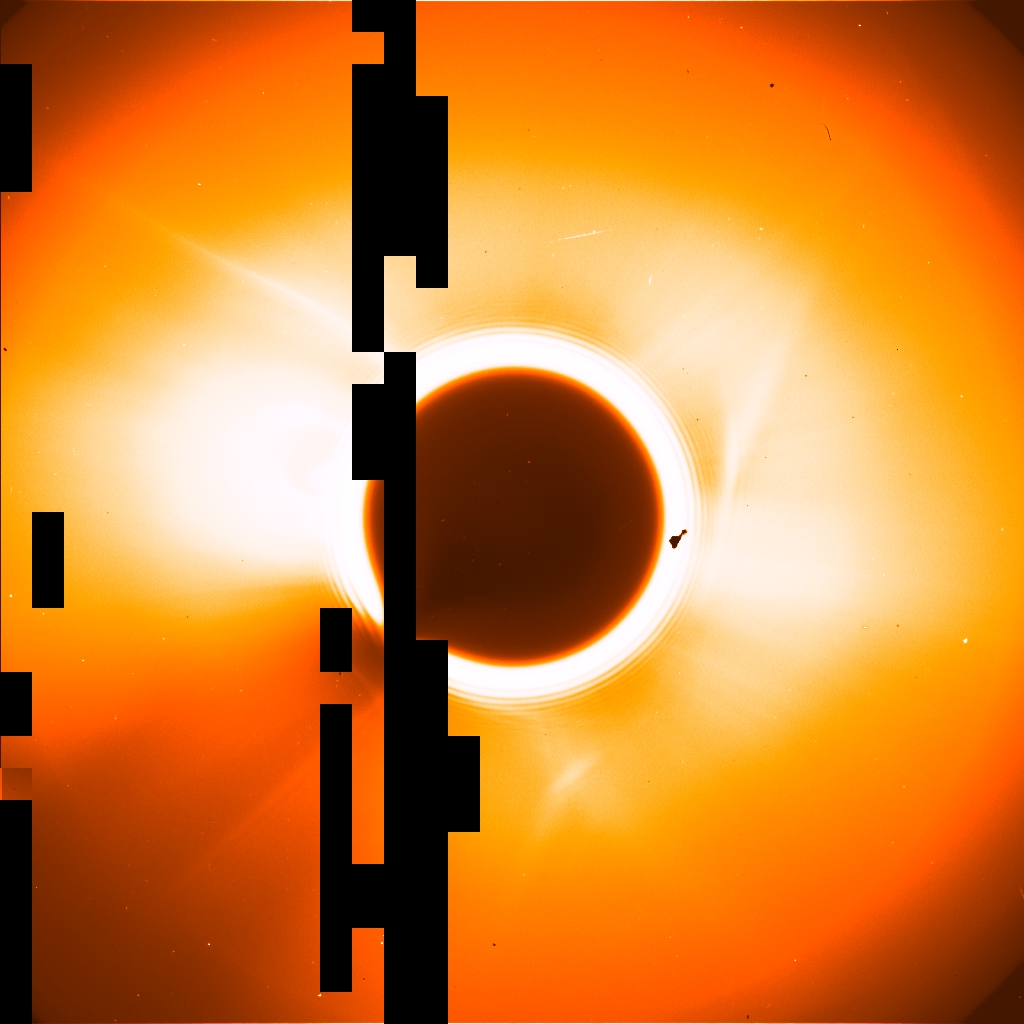}
\label{fig:C2}}
\subfloat[Subfigure 4 list of figures text][LASCO C3]{
\includegraphics[width=0.35\textwidth,scale=0.6]{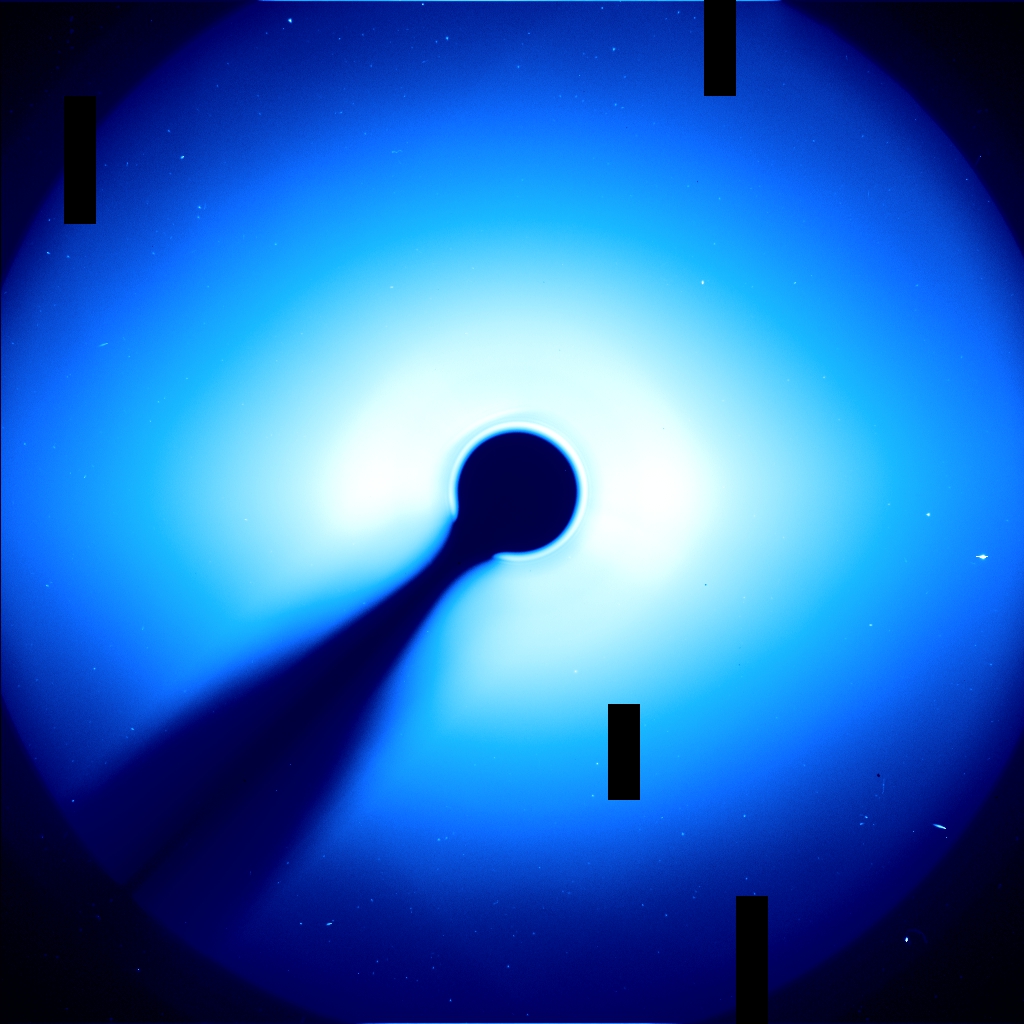}
\label{fig:C3}}
\qquad
\subfloat[Subfigure 5 list of figures text][Mercury Transit in LASCO C2]{
\includegraphics[width=0.35\textwidth,scale=0.6]{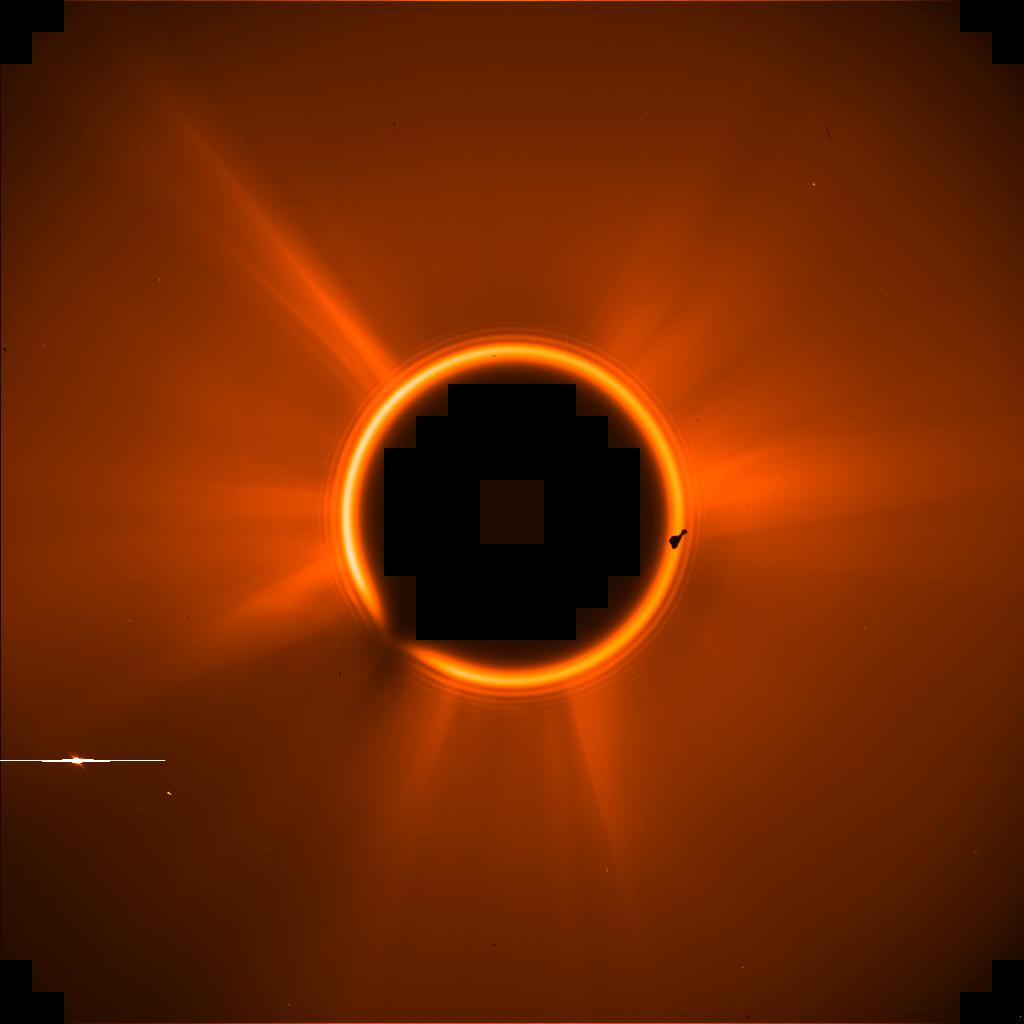}
\label{fig:MercTrans_C2}}
\subfloat[Subfigure 6 list of figures text][Mercury Transit in LASCO C3]{
\includegraphics[width=0.35\textwidth,scale=0.6]{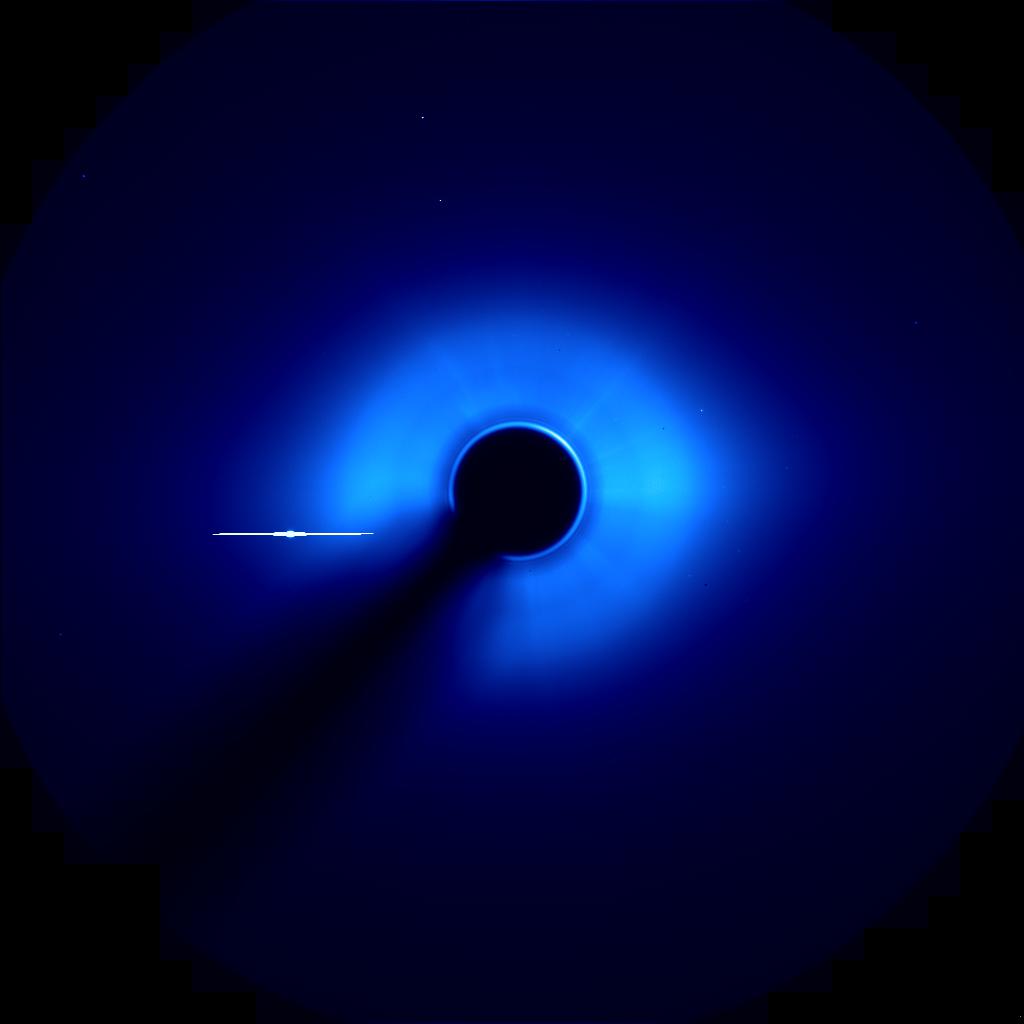}
\label{fig:MercTrans_C3}}
\caption{Examples of missing pixel data in the various SoHO products (a-d) as well as planetary transits in the LASCO coronagraphs (e-f). Telemetry errors take the form of large missing portions of the field of view, single or multiple image `holes', and `stripes'. Colors similar to those used on the NASA SoHO site have been used.}
\label{fig:holes}
\end{figure}

\begin{table}[ht]
\centering
\begin{tabular}{|l|l|l|l|l|l|l|}
\hline
Instrument & Detector & Observation & Observed Region & $\lambda($\AA$)$ & Cadence (min) & Date Range\\
\hline
MDI & MDI & LOS Mag. Fld. & Full Disk & $6768$ (Ni I) & $\sim 96$ & 1996.05.01 - 2011.04.12 \\
\hline
EIT & EIT & Intensity & Full Disk & $171$ (Fe IX/X) & $\sim 360$ & 1996.01.01 $\rightarrow$ \\
EIT & EIT & Intensity & Full Disk & $195$ (Fe XII) & $\sim 12$ & 1996.01.01 $\rightarrow$ \\
EIT & EIT & Intensity & Full Disk & $284$ (Fe XV) & $\sim 360$ & 1996.01.01 $\rightarrow$ \\
EIT & EIT & Intensity & Full Disk & $304$ (He II) & $\sim 360$ & 1996.01.01 $\rightarrow$ \\
\hline
LASCO & C2 & Intensity & Corona ($1.5 - 6~R_s$) & Visible & $\sim 20$ & 1995.12.08 $\rightarrow$ \\
LASCO & C3 & Intensity & Corona  ($3.5 - 30~R_s$) & Visible & $\sim 20$ & 1995.12.08 $\rightarrow$ \\
\hline
\end{tabular}
\caption{\label{tab:SoHO}Suite of SoHO Instruments utilized. LOS Mag. Fld. denotes the line-of-sight magnetic field, $\lambda($\AA$)$ is wavelength measured in angstroms, and $R_s$ is the Sun's radius. LASCO C1 ($1.1 - 3~R_s$) is not included in this work since it was only operational till Aug. 9, 2000.} 
\end{table} 

Stanford University's Joint Science Operation Center (JSOC) stores data from SoHO MDI, SDO HMI and AIA, and various other solar instruments. The SunPy-affiliated package DRMS enables querying of these images\cite{Glogowski2019, sunpy20}. 
All of these individual image products from JSOC are at the same processing level and are supplied in a Flexible Image Transport System (FITS) format that contains only scalar values.
The NASA Solar Data Analysis Center's (SDAC) Virtual Solar Observatory\cite{vso} (VSO) tool enables data queries from a number of individual data providers. 
SDAC's terabytes of available EIT and LASCO images are also in FITS format. 
However, the SDAC data is highly heterogeneous. Not only are there intrinsic differences among these SoHO products (e.g., individual cadence as shown in Table~\ref{tab:SoHO}), but 
there is also an irregular assortment of image file sizes and processing levels. 
These varying file sizes can correspond to different image resolutions, calibrations and orbital maneuvers, and multi-frame recordings. 
In addition, all four of the publicly available EIT products require calibration to account for instrument degradation and the presence of the burn-in caused by continuous exposure to the sun. 
SunPy's\cite{sunpy15, sunpy20} Federated Internet Data Obtainer (Fido) interface with the VSO helps to reduce the search space by enabling a query based on a priori knowledge of the appropriate file sizes but is currently limited in its time sampling capabilities. 
In addition, a few percent of the total data for each of these mission products contains artefacts, illustrated in Fig.~\ref{fig:holes}~(a-d). 
Besides holes, bright lines caused by comet transits, speckle patterns arising from cosmic rays impinging on the cameras, and spurious image artefacts, all contaminate the two coronagraphs' field of view. 
However, the dominant contribution, by far, to image contamination comes from the regularity of planetary transits that all leave a bright horizontal signature on the LASCO image products, an example of which is shown in Fig.~\ref{fig:holes}~(e-f).
Hence, preparing an ML-ready dataset from solar images, that is a dataset that can be ingested by a machine learning algorithm, remains challenging, time-consuming, and error-prone. 

In the event that a sufficiently complex machine learning algorithm would be able to extract sufficient information from the images, either to improve the state-of-the-art forecasting, or to deepen our understanding of the underlying physical processes, this would constitute an invaluable resource for the community. On the other hand, a null result would also be an important step that would highlight the present limitations of solar images for space weather prediction.

\section{Methods}\label{sec:methods} 
In this section we provide an overview of the two-step process to the Python tool (\url{https://github.com/cshneider/soho-ml-data-ready}) for producing a machine-learning-ready community standard dataset from solar images. 

\begin{figure}[h!]
\centering
\includegraphics[width=1.0\linewidth]{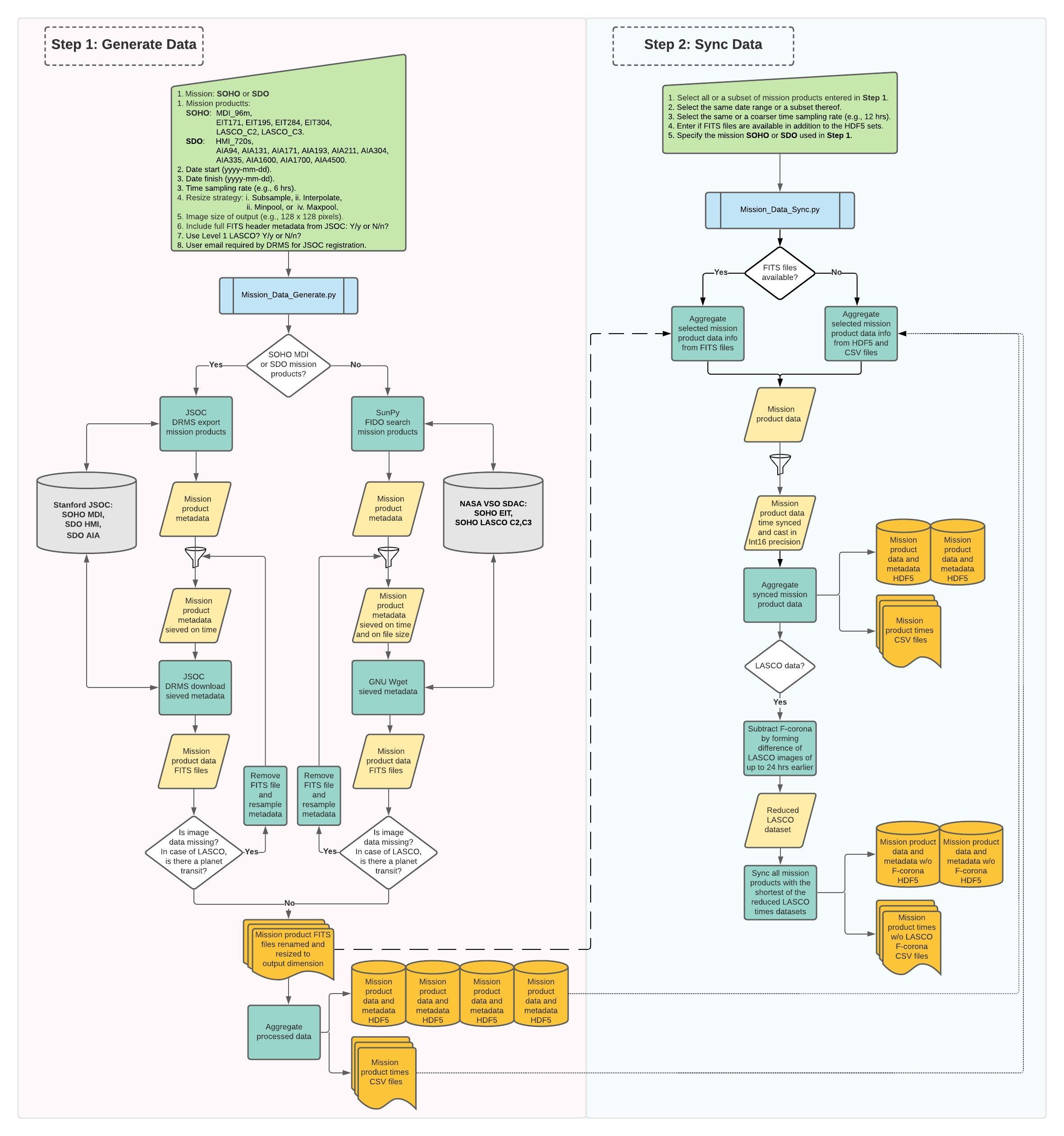}
\caption{The two flowcharts describe the two main steps of the standard dataset preparation pipeline: data generation and data synchronization. The following color scheme is used: light green denotes software options provided for the user, light blue is for the executing program, turquoise is for key main actions, gray is for external data, yellow is for flowing data, white is for a decision, and orange is for output data products.}
\label{fig:pipeline_steps1_2}
\end{figure}

\subsection{Data preparation}\label{subsec:data} 
The data preparation pipeline starts with a data generation step. This is illustrated in the leftmost flowchart of Fig.~\ref{fig:pipeline_steps1_2}.

\begin{enumerate}
    \item[] \textbf{Step 1 (data generation)}:
    \item  User inputs:
    \begin{itemize}
        \item The user selects the SoHO or SDO mission followed by a set of mission products, 
        \item Appropriate mission start and end dates are required,  
        \item A user-defined time window specifies the time sampling rate of the data in hours, 
        \item In order to be able to fit all of the data on a several GB GPU to speed up machine learning experiments, it is necessary to down-sample the native resolution of the images (i.e., (1024 x 1024) pixels for SoHO and (4096 x 4096) pixels for SDO) to a user-defined output image size. Since the information content of the magnetogram, EUV/UV/Visible, and coronagraph data products is fundamentally different, four different resize strategies are provided per code execution. For instance, one can perform three separate code runs with min-pooling on MDI, max-pooling on EIT, and sub-sampling for LASCO. In the event that the user wants to retain SoHO products that contain missing data that meets an acceptable threshold, an interpolation option is also provided for down-sampling. The definitions of these operations are as follows: i. sub-sampling selects pixels in every image row and column that are separated by a step size given by a scale factor. This scale factor is equal to the ratio of the original image's axis length to the desired image output size (e.g., 1024/128). ii. the interpolation option is a linear interpolation that shrinks the image by the scale factor, has spatial anti-aliasing which minimizes distortion artefacts known as aliasing that occur when a high-resolution image is represented at a lower resolution, and preserves the original range of values in the output image. iii-iv. min-pooling and max-pooling both use the block-reduce function from Scikit-image\cite{scikit-learn} to tile the image in blocks of size (scale factor) x (scale factor) and compute the minimum or maximum on each such block,
        \item FITS headers are down-sampled using the FITS convention with compliant SunPy keywords corresponding to the down-sampling of the respective image products,
        \item An option of including the updated FITS metadata with compliant SunPy keywords for JSOC products is also provided
        but it adds to the processing time as the server needs to first prepare the requested files,
        \item 
        As an alternative to using raw, uncalibrated LASCO Level-0.5 data, the option of using Level-1.0 LASCO images is provided. LASCO Level-1.0 data products have the following corrections applied: i. corrections for bias, and calibration to units of mean solar brightness, ii. correction to account for vignetting, particularly around the occulting disk and pylon, iii. warping image to yield flat geometry for entire field of view, iv. up to several minutes clock correction, v. updates to metadata parameters such as the Sun-center location and exposure time. Furthermore, since 2003, the SOHO spacecraft performs 180-degree flips approximately every three months to account for an antenna issue; the Level-1.0 data product provides precise corrections for these roll periods in the metadata, though does not rectify images. 
        In our opinion, these corrections are not needed for input to an ML system for three reasons: 
        a. image flipping is generally an invariant operation for an ML system   
        and is used for image augmentation to generate more examples for ML systems; 
        b. image warping may interfere with the integrity of the raw data; and c. using the LASCO Level-1.0 data in calibrated physical units provides no advantage over the raw Level-0.5 data which has units of photon counts (DN). 
        In regard to calibrated EIT images, it is anticipated that Level-1.0 EIT images will become publicly available for querying via an online interface hosted by ESA later this year,
        \item For use of the DRMS system, JSOC requires that the user preregister their email address at \url{http://jsoc.stanford.edu/ajax/register_email.html}.
      \end{itemize}   
    \item Databases and queries:
        \begin{itemize}
        \item With the user selection of SoHO MDI or SDO mission products, Stanford's JSOC database is queried using DRMS export rather then DRMS query which also lists `ghost' files that are of size zero. The following JSOC series are used: i. \texttt{mdi.fd\_M\_96m\_lev182} for SoHO MDI at a natural cadence of 96 minutes, ii. \texttt{hmi.M\_720s} for SDO HMI at a natural cadence of every 720~seconds, iii. \texttt{aia.lev1\_euv\_12s} for SDO EUV AIA (94, 131, 171, 193, 211, 304, 335~\AA) at a natural cadence of every 12 seconds, iv. \texttt{aia.lev1\_uv\_24s} for SDO UV AIA (1600 and 1700~\AA) at a natural cadence of 24 seconds, v. \texttt{aia.lev1\_vis\_1h} for SDO visible AIA (4500~\AA) at a natural cadence of 1 hour.  
        For SDO AIA, both `images' and `spikes' (for calibration) are available and so `images' are explicitly specified. DRMS download is used to retrieve the DRMS exported data, 
        \item With the user selection of SoHO EIT, SoHo LASCO~C2, and/or LASCO~C3, SunPy's Fido is used to query the NASA VSO SDAC database. GNU's Wget\cite{gnuwget} is used to fetch the images since it avoids timing out on large datasets, 
        \item An internal step size of 60 days is used to produce the initial queries for all SoHO mission and SDO mission products and it respects SDAC's $10$k results quota per search as well as DRMS' $100$ GB export limit.
        \end{itemize}
    \item Reduction of data search space:
        \begin{itemize}
        \item The key difference between the two query tracks in terms of metadata sieving is that JSOC products need only to be sieved on time (i.e., user-defined time sampling rate or cadence) whereas SDAC products need to be sieved on time and on file size. 
        Sieving is performed by computing a dynamic set of indices which is updated each time that a candidate file, that has been locally downloaded, is found to have missing data or to have the wrong shape or to have planetary transits in the case of LASCO products.
        The actual download proceeds from singleton download requests which are sequentially issued based on these indices. 
        Selecting the proper sized file sizes from the database, a priori, offers a major performance advantage over having to locally download and individually examine FITS headers. 
        Local downloading of files is consequently held off until the cleaning step of the pipeline which utilizes Astropy\cite{astropy:2013,astropy:2018} to access the FITS data and FITS keywords,
        \item The planet transit filter works by applying the Probabilistic Hough Transform on an edge-filtered image, obtained from using the Canny algorithm, to specifically detect only horizontal lines. Scikit-image is used for both algorithms,  
        \item Native SoHO image data corresponds to a resolution of (1024 x 1024) pixels with the exception of EIT~195 which is also frequently available in (512 x 512) pixel resolution. These product resolutions correspond to specific file sizes as shown in Table~\ref{tab:datasizes}. In order to efficiently reduce the VSO query size for these proper data from the very beginning, SunPy's Fido `search' function is used to return a `QueryResponseBlock' dictionary object which contains the keyword `size' in addition to other keywords corresponding to columns 1-5 and 7 of Table~\ref{tab:SoHO} as well as the `fileid' url pointer for file download. 
        In just a handful of cases, the file size on its own is not sufficient to guarantee the proper data type which necessitates `naxis=2' in the FITS header to be true in order to have a 2D image as opposed to a 3D data cube.
        \end{itemize}
    \item Mission product FITS with data and metadata cubes and corresponding times:
        \begin{itemize}
        \item In the event of an unlikely interruption in downloading which crashes the program, the user can restart on the last FITS file date downloaded, 
        \item FITS files are first generated from all data satisfying user specifications, 
        \item The data from these FITS files is then aggregated and delivered in compressed Hierarchical Data Format version 5\cite{collette_python_hdf5_2014,hdf5} (HDF5) data cubes per data product specified. Accompanying each of these data cubes are comma-separated values (CSV) files containing the time axis points of the respective data cube. These data cubes are now ready for input to the second step of the software pipeline which is data synchronization. 
        \end{itemize}
\end{enumerate}

A major problem in using solar images for machine learning is that, as shown in Table~\ref{tab:SoHO}, each image comes with its own time cadence. In our Python tool, the user-defined time sampling rate fixes the cadence of all products to the same rate and syncing makes the user-selected set of SoHO products all fall within the same time window. 
This is illustrated in the rightmost flowchart in Fig.~\ref{fig:pipeline_steps1_2}. This operation further reduces the volume of data to a more manageable temporal resolution.

\begin{enumerate}
  \item[] \textbf{Step 2 (time synchronization):}
  
    \item User inputs
        \begin{itemize}
        \item The user selects the mission entered in Step 1 together with all or a subset of the mission products and all or a subset of the date range, as well as the same or coarser time sampling rate sub-sampled by an integer multiple. 
        The user also has the choice of combining images of the same output dimension but obtained with different resize strategies across the different products,
        \item  The user is presented with two options: i. if the user has simply downloaded the HDF5 cubes and CSV files without locally generating them, then the data from the HDF5 cubes will be taken up along with the corresponding times from the accompanying CSV file for each product specified, ii. if the user has run the Python tool and obtained both the HDF5 data cubes along with the corresponding FITS files then the FITS files will be used to obtain both the data and the times since the time stamps are written into the names of the FITS files themselves, 
        In the case of FITS files being present, the times are directly read from the names of the FITS files because this is more robust against any interruption in the data generation step,
        \item The sync algorithm then finds the product with the shortest time list and moves along its time axis checking for overlap with the times of the other products. The overlap time is defined as +/- half the time window. The closest times from all products to that time are taken and if equal times are found for a product then the first such time is always selected. If a matching time for a product for a particular point on the time axis is not found, all other product times obtained for that time axis point are discarded. Therefore, it follows that the more products specified, the harder it is for a time axis point to be shared among all the products,
        \item The outputs are synced HDF5 cubes 
        in a numeric format suitable for efficient GPU utilization
        and corresponding synced times per product as CSV files,
        \item If a LASCO product is present, the F-corona is subtracted by taking the time difference between the given LASCO image with itself up to 24 hours earlier. This results in a reduced LASCO image set which is then synced with the rest of the selected products,
        \item This new reduced set is then output as the second set of HDF5 and CSV files.
        \end{itemize}
\end{enumerate}

Cleaning and processing from the first step and syncing from the second step together yield a machine-learning-ready standardized dataset.

\section{Data Records} \label{sec:datarecords}

\begin{table}[ht]
\centering
\begin{tabular}{|l|l|l|l|l|}
\hline
SoHO product & File size (KB) & Resolution (pixels) & Calibration level & Provider \\
\hline
MDI & $\sim$ 1400-1600 & (1024 x 1024) & L1.8.2 & JSOC \\ 
\hline
MDI & 4115 & (1024 x 1024) & L1.5, L1.8.2 & Removed from SDAC \\
\hline
EIT 171 & 2059 & (1024 x 1024) & L0.5 & SDAC \\
\hline
EIT 195 & 2059 or 523 & (1024 x 1024) or (512 x 512) & L0.5 & SDAC \\
\hline
EIT 284 & 2059 & (1024 x 1024) & L0.5 & SDAC \\
\hline
EIT 304 & 2059 & (1024 x 1024) & L0.5 & SDAC \\
\hline
LASCO C2 & $\sim2100$, 4106 & (1024 x 1024) & L0.5, L1.0 & SDAC \\ 
\hline
LASCO C3 & $\sim2100$, 4106 & (1024 x 1024) & L0.5, L1.0 & SDAC \\ 
\hline
\end{tabular}
\caption{\label{tab:datasizes} 
Proper file sizes identified for all available calibration levels of SoHO products from the SDAC and JSOC data providers with their respective resolutions.
LASCO file sizes are given as approximate due to some minor size variations noted over the years.
File size is not relevant for calibrated MDI images obtained from JSOC.
It is anticipated that Level-1.0 EIT images will become publicly available for querying via an online interface hosted by ESA later this year. 
}
\end{table} 


\begin{table}[ht!] 
\centering
\begin{tabular}{|l|c|c|c|c|l|}
\hline
SoHO product & No. FITS files & HDF5 cube size (MB) & No. excluded files with holes (\& planetary transits)  \\ 
\hline
MDI & 15456 & 561 & 66  \\ 
\hline
EIT 171 & 10234  & 187 & 1045 \\ 
\hline
EIT 195 & 14439 & 254 & 1391  \\ 
\hline
EIT 284 & 9601 & 151 & 895  \\ 
\hline
EIT 304 & 9439 & 158 & 936 \\ 
\hline
LASCO C2 & 15383 & 329 & 489 (\& 7511) \\ 
\hline
LASCO C3 & 13354 & 277 & 699 (\& 23901) \\ 
\hline
\end{tabular}
\caption{\label{tab:runtime} Individual SoHO products queried at a cadence of 6 hours and time span of Jan.~1,~1999 - Dec.~31,~2010. Image output dimensions are (128 x 128) pixels. The number of images that have been excluded per image product due to holes and, in the case of LASCO, planetary transits is also provided.
}
\end{table}

\begin{table}[ht]
\centering
\begin{tabular}{|l|c|c|c|c|c|}
\hline
SoHO experiment & Total sync output & Train & Validation & Test & Total sync used \\
\hline
MDI only & 15455 & 9090 & 1269 & 1227 & 11586 \\
\hline
MDI, EIT~195, LASCO~C2 & 12299 & 7384 & 953 & 943 & 9280 \\
\hline
MDI, EIT~195, LASCO~C2~$\triangle$ & 12132 & 7292 & 940 & 926 & 9158 \\
\hline
MDI, EIT~171,~195,~284,~304, LASCO~C2,~C3 & 4348 & 2369 & 275 & 347 & 2991 \\
\hline
MDI, EIT~171,~195,~284,~304, LASCO~C2~$\triangle$, ~C3~$\triangle$ & 3360 & 1833 & 195 & 268 & 2296 \\
\hline
\end{tabular}
\caption{\label{tab:sixhrcad} Five SoHO experiments selected to be synced with a cadence of 6 hours and time span of Jan.~1,~1999 to Dec.~31,~2010. The total synced output comes from the second step of the pipeline. 
For the three product and seven product experiments, the subtraction of the F-corona by time differences of up to 24 hours prior to a LASCO image and itself are used and represented by `$\triangle$' after C2,C3.
The train, validation, and test data split is formed by applying a progressive rolling indices scheme by successive year. Image dimensions are the same as reported in Table~\ref{tab:runtime}.
Note that due to the synchronization step the MDI-only experiment contains one less file than in Table \ref{tab:runtime}.
} 
\end{table} 

The input data and provenance corresponding to the suite of SoHO instruments described in Table~\ref{tab:SoHO} is provided by Table~\ref{tab:datasizes}. 
The datasets analysed during the current study, consisting of data plus metadata serialized with JSON, are available as compressed HDF5 data cubes through SURF at (\url{https://surfdrive.surf.nl/files/index.php/s/NYHm1b9hOKMMcw0}) and are all contained in the ``\text{Mission\_ML\_Pipeline}'' directory. 


The HDF5 data cubes come with corresponding time stamps per image slice provided by the accompanying CSV files. 
Time stamps are provided as a separate CSV file primarily because a CSV file provides ease-of-access for a beginning user and enables to quickly determine the corresponding number of images in the respective data cube.
Each slice of the data cube has one fixed, user-specified output image size which, in this case, is (128 x 128) pixels obtained from the sub-sampling resize strategy. 
An image output size of (128 x 128) pixels is used for all image products in this paper because all of the data for a given experiment can be efficiently loaded into GPU memory.
The original size of the images is (1024 x 1024) pixels with the exception of EIT~195 which also comes in (512 x 512) sizes. 

The first set of data cubes encompass all seven SoHO products as shown in Table~\ref{tab:runtime} and has a temporal cadence of six hours and time span of Jan.~1,~1999 - Dec.~31,~2010. 
The second set of data cubes are the five synced experiments with the same time span and temporal cadence of six hours and separately comprised of one, three, and seven image products as shown in Table~\ref{tab:sixhrcad}. 
Within the three and seven product experiments, the effect of both including and subtracting the F-corona from the LASCO products is investigated.
Explicitly, the experiments are:
i. MDI-only, ii-iii. MDI, EIT~195, LASCO~C2 (minus F-corona), iv-v. all seven SoHO products (minus F-corona). 
With more products there is a higher probability that some of them are unavailable at the specified times, so that fewer synced times are available.


The sub-directories consist of HDF5 data cubes along with their respective CSV files containing all of the time stamps and are as follows: 
\begin{itemize}
    \item ``19990101\_20101231\_6\_128\_7\_products'' are the generated data cubes for all seven products individually, 
    \item ``19990101\_20101231\_6\_128\_MDI\_self\_synced'' is the self synced MDI data cube,
    \item ``19990101\_20101231\_6\_128\_3\_products\_synced'' are the synced data cubes corresponding to MDI, EIT~195, and LASCO~C2. A further sub-directory ``Fcorona\_subtracted\_3products'' contains the corresponding F-corona subtracted products, 
    \item ``19990101\_20101231\_6\_128\_7\_products\_synced'' are the synced data cubes of all seven SoHO products. A further sub-directory ``Fcorona\_subtracted\_7products'' contains the corresponding F-corona subtracted products.
\end{itemize}
These datasets were derived from the following public domain resources: VSO (\url{https://sdac.virtualsolar.org/cgi/search}) and JSOC (\url{http://jsoc.stanford.edu/MDI/MDI_Magnetograms.html}). 
The example datasets are 6.6 GB in total. 

\section{Technical Validation} \label{sec:techvalid}



\begin{figure}[h!]
\centering
\includegraphics[width=0.85\linewidth]{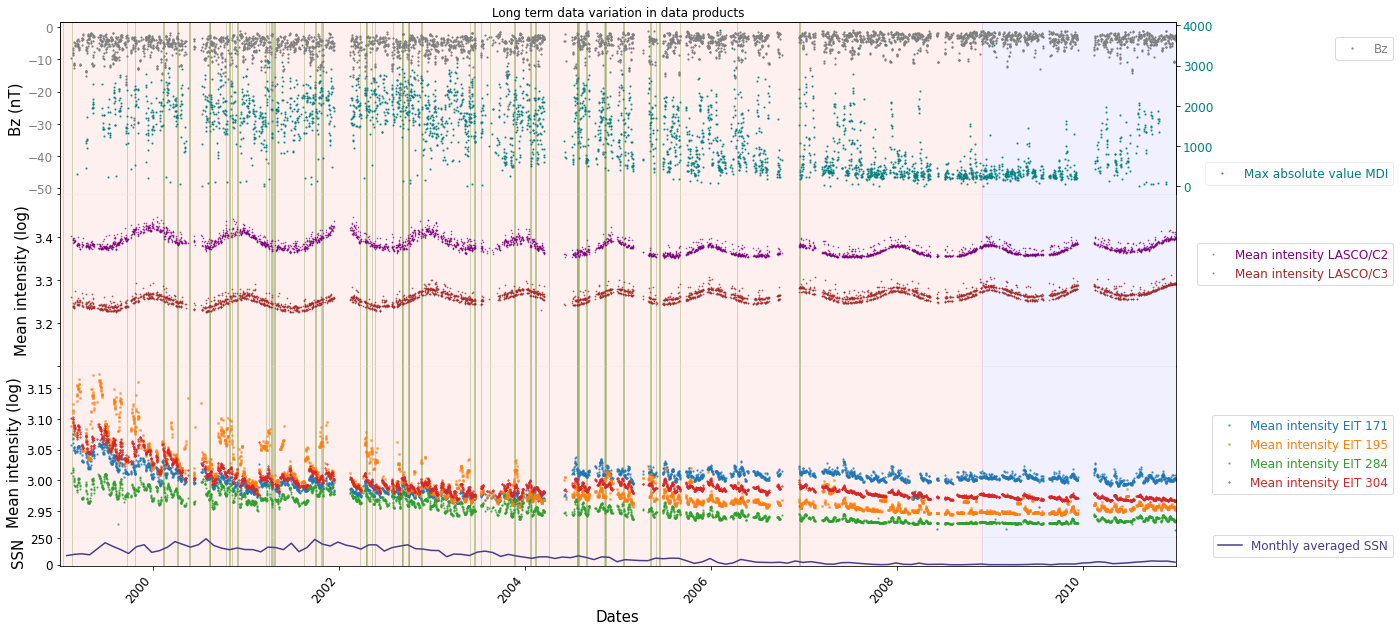}
\caption{Long term variation (1999 - 2010) of the synced products. Shown are the LASCO images with F-corona subtracted for mean intensity of the solar disc for EIT EUV (blue, red, orange and green curves) in log scale, LASCO white-light (brown and purple curves) in log scale, maximum of absolute value of MDI magnetogram (light blue curve) and $B_z$ values at L1 (gray curves). The light pink and blue backgrounds indicate the data time overlap with solar cycles 23 and 24. The solar cycle is also evident in the bottom plot with the monthly averaged sunspot number (SSN) for the duration of the dataset. The vertical olive-green lines corresponds to large geomagnetic storms with Dst indices smaller than -100. 
}
\label{fig:intensities}
\end{figure}

Having a standardized dataset facilitates data exploration including, for example, studies of correlation between different channels. 
Figure~\ref{fig:intensities} shows a comparison of mean intensity variation across all synced SoHO products from 1999 - 2010. 
The long-term variation in the mean value of the intensity of the solar disk for EIT extreme ultraviolet (EUV) images at wavelengths of 171, 195, 284, 304 {\AA} show an interesting downward trend corresponding to the CCD degradation over time. 
Also in this trend are regular clean-up operations which involve reheating the detector to account for the burn-in from the sun which leave `break-like' signatures in the spectrum. 
The wiggles on the spectrum have a periodicity of roughly 27 days and correspond to active regions. 
Less readily visible is the growth of mean intensity with increasing solar cycle 23. 
One of the most interesting peaks seen in all the EUV channels occurs on 
October 27, 2003 
corresponding to the Halloween storm of 2003 \cite{halloween_2003, halloween_Nat} a combination of the bright active regions and the excessive particle events that `blinded' the camera repeatedly over those few days. 
The long term variation of the LASCO~C2 and LASCO~C3 both show a periodic variation in the mean intensity. 
The cyclical broadening of the peak is the puffing up of the F-corona every half year due to the apparent size of the sun getting larger than the fixed occulter and then getting smaller again. This cyclical variation with the Sun–SoHO distance was also found in previous studies\cite{tempvar}. 
The overall amplitude in the LASCO images follows the solar cycle and the tiny wiggles, as for the EIT images, correspond to Carrington rotations and harmonics thereof.
The peaks in the mean intensity of the images for LASCO~C2 and LASCO~C3 are partly attributed to observational artefacts and to CMEs in LASCO field of view. The signature of the aforementioned Halloween storm is also visible in the mean value of the LASCO images.  
The maximum of the absolute values of the magnetic field measurements by MDI best represents the solar cycle; getting weaker approaching the minimum of solar cycle 23 in the later part of 2008 and rising up again, marking the beginning of solar cycle 24. A visual comparison can be seen between the observations of the MDI measurements with the monthly averaged sunspot number (SSN), marking the maxima and the declining phase of the solar cycle 23 and beginning of the solar cycle 24. 
The gaps in the data products are due to the prepossessing step where the planetary transits are filtered out from the entire dataset. 

Accompanying the SoHO product temporal profiles, dips in the north-south component of the interplanetary magnetic field (IMF) $B_{z}(t)$, as measured at Lagrange point one (L1), signify the geomagnetic storms.
The stark difference in the geomagnetic activity at different phases of the solar cycle can be clearly seen. The dips are numerous during the maximum of the solar cycle 23. In contrast, the rising phase of the solar cycle does not show significant dips in the $B_z$ measurements at L1.
As seen from Figure~\ref{fig:intensities}, pre-processing of the data is clearly non-trivial and additional steps downstream of our pipeline would be needed to account for instrument degradation and trends in the data that are entangled with solar cycle variability. 


\section{Usage Notes} \label{sec:usagenotes}


Along with the dataset, we provide and discuss results for a simple machine learning model that predicts the onset of a geomagnetic storm. This is intended as a baseline model that can be used in future studies to assess the skill 
of more complex models. 

\begin{table}[h!]
\centering
\begin{tabular}{|l | c | c c c c | c | c |}



\hline
Product used for Gaussian Naive Bayes (GNB) & $\%$  & TP & FP & FN & TN & TSS & MCC \\ 
\hline 
MDI (max, min, std) & $25$ & 212 & 336 & 85 & 594 & 0.35 & 0.30    \\ 
MDI (min, std, frac) & $20$ & 171 & 392 & 67 & 597 & 0.32 & 0.26    \\ 
MDI (min, std, frac) & $15$ & 141 & 392 & 50 & 644 & 0.36 & 0.26    \\ 
MDI (min, std, frac) & $10$ & 110 & 400 & 36 & 681 & 0.38 & 0.25    \\ 
MDI (min, std, frac) & $5$ & 35 & 459 & 5 & 728 & 0.49 & 0.18    \\ 
\hline          
MDI (std), EIT~195 (frac),  LASCO~C2 (0) & $25$  & 153 & 252 & 67 & 471 & 0.35 & 0.30 \\
MDI (min), EIT~195 (0),  LASCO~C2 (0) & $20$ & 131 & 338 & 45 & 429 &  0.30 & 0.24  \\
MDI (std), EIT~195 (frac), LASCO~C2 (0) & $15$ & 96 & 273 & 47 & 527 & 0.33 & 0.24  \\
MDI (std), EIT~195 (frac), LASCO~C2 (0) & $10$ & 77 & 298 & 36 & 532 &  0.32 & 0.21  \\
MDI (std), EIT~195 (min), LASCO~C2 (frac) & $5$ & 32 & 441 & 2 & 468 & 0.46 & 0.17  \\ 
\hline
MDI (frac), EIT~195 (mean), LASCO~C2~$\triangle$ (frac) & $25$  & 159 & 378 & 53 & 336 & 0.22 & 0.19 \\ 
MDI (frac), EIT~195 (std), LASCO~C2~$\triangle$ (frac) & $20$ & 119 & 333 & 51 & 423 & 0.26 & 0.20  \\
MDI (frac), EIT~195 (mean), LASCO~C2~$\triangle$ (frac) & $15$ & 117 & 448 & 22 & 339 & 0.27 & 0.20  \\ 
MDI (frac), EIT~195 (mean), LASCO~C2~$\triangle$ (frac) & $10$ & 92 & 418 & 17 & 399 &  0.33 & 0.22  \\
MDI (frac), EIT~195 (std), LASCO~C2~$\triangle$ (frac) & $5$ & 30 & 401 & 3 & 492 & 0.46 & 0.17  \\ 
\hline
MDI (std), 195 (0), 171 (max), 284 (std), 304 (std), C2 (max), C3 (min) & $25$ & 57 & 72 & 29 & 189 & 0.39 & 0.35 \\
MDI (frac), 195 (mean), 171 (std), 284 (std), 304 (max), C2 (max), C3 (std) & $20$  & 39 & 66 & 23 & 219 & 0.40 & 0.33 \\
MDI (min), 195 (mean), 171 (max), 284 (max), 304 (std), C2 (max), C3 (frac) & $15$  & 38 & 101 & 10 & 198 & 0.45 & 0.32 \\
MDI (frac), 195 (mean), 171 (std), 284 (std), 304 (std), C2 (max), C3 (0) & $10$  & 27 & 63 & 12 & 245 & 0.49 & 0.35 \\
MDI (frac), 195 (0), 171 (max), 284 (min), 304 (std), C2 (max), C3 (mean) & $5$  & 9 & 45 & 2 & 291 & 0.68 & 0.33 \\
\hline
MDI (frac), 195 (std), 171 (max), 284 (min), 304 (std), C2~$\triangle$ (mean), C3~$\triangle$ (mean) & $25$ & 26 & 14 & 39 & 189 & 0.33 & 0.40 \\
MDI (max), 195 (mean), 171 (min), 284 (0), 304 (std), C2~$\triangle$ (max), C3~$\triangle$ (frac) & $20$ & 33 & 91 & 13 & 131 & 0.31 & 0.23 \\ 
MDI (max), 195 (mean), 171 (max), 284 (mean), 304 (std), C2$\triangle$ (std), C3~$\triangle$ (0) & $15$  & 28 & 74 & 8 & 158 & 0.46 & 0.32 \\
MDI (max), 195 (std), 171 (0), 284 (mean), 304 (std), C2~$\triangle$ (min), C3~$\triangle$ (min) & $10$  & 25 & 84 & 4 & 155 & 0.51 & 0.32 \\
MDI (max), 195 (frac), 171 (min), 284 (mean), 304 (max), C2~$\triangle$ (min), C3~$\triangle$ (mean) & $5$  & 8 & 28 & 1 & 231 & 0.78 & 0.41 \\
\hline
Product used for Convolutional Neural Network (CNN) & $\%$  & TP & FP & FN & TN & TSS & MCC \\
\hline
MDI only & $25$ & 15 & 13 & 282 & 917 & 0.04 & 0.10   \\
MDI, EIT~195, LASCO~C2 & $25$ & 29 & 25 & 191 & 698 & 0.10 & 0.18 \\
MDI, EIT~195, LASCO~C2~$\triangle$ & $25$ & 53 & 145 & 159 & 569 & 0.05 & 0.05 \\
MDI, EIT~195, EIT~171, EIT~284, EIT~304, LASCO~C2, LASCO~C3  & $25$  & 27 & 20 & 59 & 241 & 0.30 & 0.24\\
MDI, EIT~195, EIT~171, EIT~284, EIT~304, LASCO~C2~$\triangle$, LASCO~C3~$\triangle$ & $25$ & 20 & 26 & 45 & 177 & 0.18 & 0.20 \\

\hline
\end{tabular}
\caption{\label{tab:results_GNB}
Results from the five SoHO experiments obtained for two categories of models for binary classification: Gaussian Naive Bayes (GNB) (first five cells) and from a three-layer convolutional neural network (CNN) (final cell). Each model is evaluated at a percentile $(25,20,15,10,5)\%$ of $B_z$ values from the MDI-only training set. These percentiles correspond to $B_z$ values of -6.28, -6.95, -7.76, -8.92, and -11.61~nT, respectively. The best performing CNN models for each of the five experiments are shown. The `$\triangle$' denotes F-corona subtraction obtained by taking the time difference between the given LASCO image with itself up to 24 hours earlier. For the GNB category, the best performing models are shown with their corresponding features per SoHO product: `max' is the maximum, `min' is the minimum, `mean' is the mean, `std' is the standard deviation, `frac.' is the Minkowski–Bouligand fractal dimension, and `0' denotes feature absence. Per model, the True Skill Score (TSS), at $50\%$, and Matthews Correlation Coefficient (MCC) are given with corresponding true positives (TP), false positives (FP), false negatives (FN), and true negatives (TN).}
\end{table}

\subsection{Example application of SoHO dataset to space weather use case}

SoHO is a cooperative mission between NASA and ESA and was the first space-based telescope to serve as an early warning system for space weather. SoHO was launched on December 2, 1995 and remains operative after 25 years of service. It has revolutionized space weather forecasting capabilities by providing several days notice of Earth-directed disturbances such as Coronal Mass Ejections (CMEs).
Space weather refers to conditions on the sun, in the solar wind, magnetosphere, ionosphere, and thermosphere, that can influence the performance and reliability of space-borne and ground-based technological systems and can endanger human life or health\cite{defSW}. 

In this section we discuss an example application of the dataset to a challenging space weather problem, namely the forecast of $B_{z}$ at L1, in a two-day time window which spans $3 - 5$ days ahead of the image data. 
This is physically justified since the mean bulk speed\cite{cme_speed} of CMEs is $400$ km/s which yields an average propagation time from Sun to Earth of around three days. 
Even the slow solar wind from the equatorial regions of the Sun has a mean speed\cite{solarwind} of $400$ km/s.
It is known that that coronal brightness alone has no relationship with $B_{z}$, as recently re-affirmed using two full solar cycles of LASCO~C2 data\cite{battams20}.  

We note that the images contain the full solar disk, so information about events affecting $B_z$ might be present before and after an image that is assigned a positive event truth label. This can lead to a solar image with an active region toward the left solar limb being labeled a negative event since its effect has not yet arrived at L1. 
In order to simplify the problem, the prediction of $B_{z}$ is treated as a binary classification problem, with the target being defined as the difference between the minimum of $B_{z}$ and the mean of $B_{z}$ 
in each two-day window. Values of $B_{z}$ are obtained from the NASA OMNI low resolution (LRO) dataset (\url{https://spdf.gsfc.nasa.gov/pub/data/omni/low_res_omni}) which provides hourly averaged values. The mean is subtracted in order to 
have some variation around the minimum value of $B_{z}$ encountered during geomagnetic disturbances since these values tend to dominate the two-day window. 
The data is selected to span the years 1999 to 2010 inclusive. This is because the intrinsic image quality from 1996 - 1998 is not too reliable. 

Five experiments are performed using the following image products: i. MDI-only, ii-iii. MDI, EIT~195, LASCO~C2 (minus F-corona), iv-v. all seven SoHO products (minus F-corona). 
For each experiment the datasets are synced.
In order to prevent temporal correlation between the training, validation and test sets, a rolling split was applied to the data. For 1999, Feb.- Aug. is used for the training set, Oct. for the validation set, and Nov. for the test set. For each subsequent year, all three of these set partitions are advanced by one month forward so that for 2010, as an example, it is Jan.-July for training, Sept. for validation, and Oct. for the test set. The other months are not used. 
The final datasets for each experiment are given in Table~\ref{tab:sixhrcad} and approximately yield an 80-10-10 split for the training, validation, and test sets.
Each of the experiments' test sets is normalized by the maximum of the absolute value of the respective training sets. 
The corresponding truth labels for each experiment are further assigned a binary value determined by setting the threshold $p$ at five percentiles (25,20,15,10,5)\% corresponding to $B_z$ values of -6.28, -6.95, -7.76, -8.92, and -11.61~nT, respectively. 
This causes a class imbalance at each threshold which requires the use of class weights for training both the CNN and GNB models to perform binary classification.

\subsubsection{Machine learning approach}\label{subsubsec:CNN}
\begin{figure}[h!]
\centering
\includegraphics[width=0.6\linewidth]{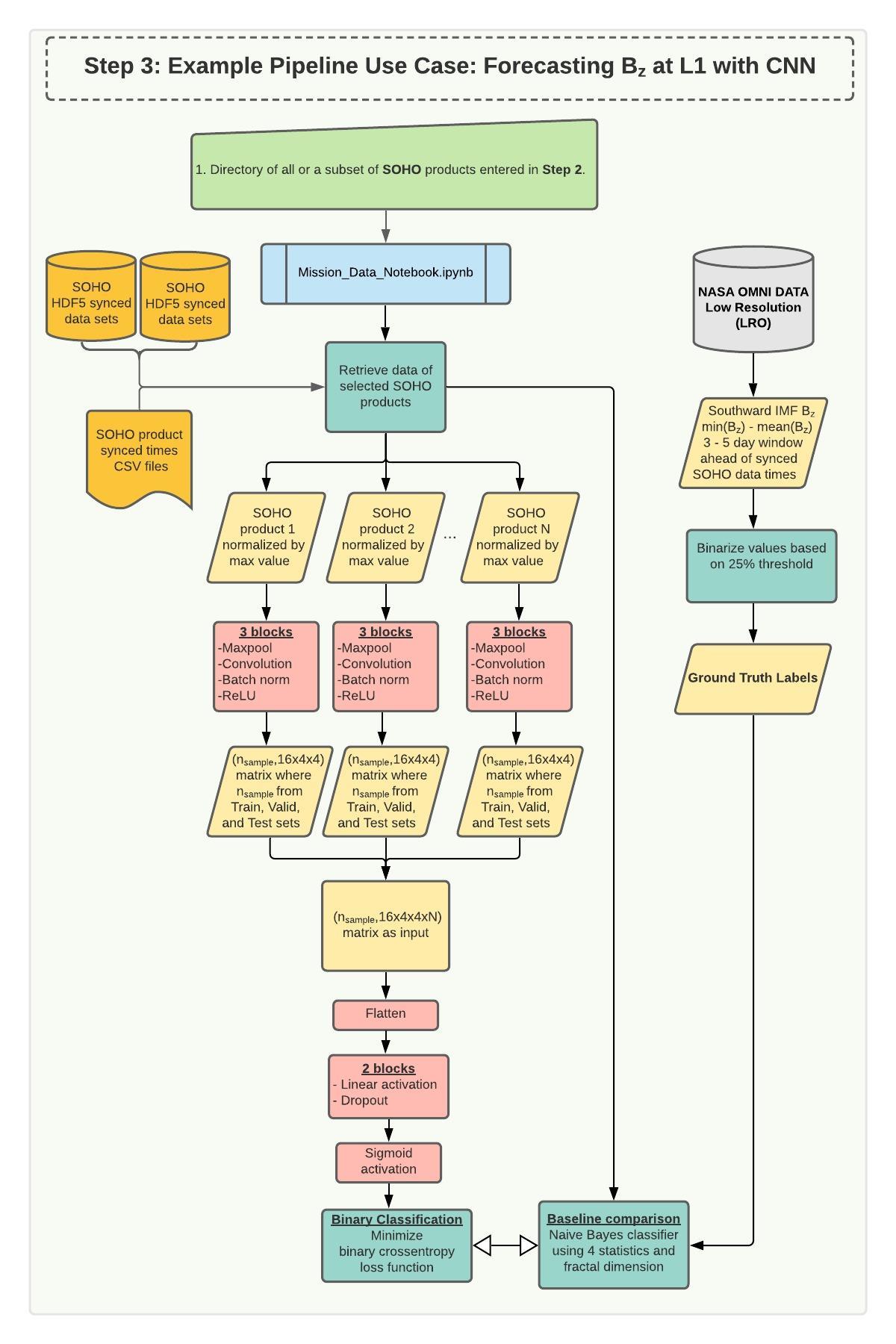} 
\caption{The flowchart describes the steps of applying a deep CNN network for classifying the occurrence of a geomagnetic storm using $B_z$ as an example proxy. This prediction is then compared with a classical baseline. 
In the box titled `Baseline comparison', the four statistics refer to the max, min, mean, and standard deviation and the fractal dimension is the Minkowski–Bouligand fractal dimension. 
Three block iterations of successive max-pooling (2 x 2), convolution (3 x 3), batch normalization and ReLU activation is applied on the (128 x 128) pixel synced images. 
The convolutional layers are each intialized using the Glorot uniform initializer, also called Xavier uniform initializer. 
After the flattened layer containing the aggregate normalized features from all products come three fully connected layers. Batch normalization is first applied, followed by two iterative blocks of successive linear activation and dropout. Linear activation consists of 128 activation units in the first block and 32 units in the second. A dropout rate of 60\% is used as a regularizer to reduce over-fitting. 
The final single activation unit is given by the sigmoid function which assigns binary values to the preceding probabilities. It is at this stage only that a single bias term is introduced. A binary cross entropy loss function is used together with a SGD optimizer with a fixed learning rate of 0.01. A convergence is achieved within 100 epochs with early stopping enabled triggered by the validation loss. 
Color scheme is same as in Fig.~\ref{fig:pipeline_steps1_2} with the addition of pink which denotes key stages in the CNN architecture. 
} 
\label{fig:pipeline_step3}
\end{figure}

We apply a three-layer CNN, implemented with TensorFlow\cite{tf15}, to each of the five experiments. 
The SoHO products used in each of these experiments constitute the input channels.
These channels are input in parallel with their individual dense or fully connected layers eventually concatenated as depicted in Fig.~\ref{fig:pipeline_step3}. 
Due to the application of a down-sampling strategy on the images, the full data cubes can fit into a several GB GPU memory with the final down-sampled sizes constrained by how many image products will be run in parallel.
In our case, sub-sampling to (128 x 128) pixels fits into the GPU for all five experiments.
This parallel approach is chosen over the traditional, stacking approach, in that the integrity of the individual product's features may be preserved instead of merging features across different products which will confound the individual product's contribution.
Three successive block iterations consisting of max-pooling, convolution, batch normalization and a rectified linear unit (ReLU) activation are applied on each parallel network. After the above mentioned concatenation, two successive blocks of linear activation and dropout, which acts as a regularizing term to prevent over-fitting, are applied. 
The final single activation unit is given by the sigmoid function which assigns binary values to the preceding probabilities. A binary cross entropy loss function is used together with a stochastic gradient descent (SGD) optimizer. 
A convergence is achieved within 100 epochs with early stopping enabled triggered by the validation loss.
The total number of model parameters is $\sim$62K, $\sim$178K, and $\sim$409K for the one, three, and seven product experiments, respectively.  
The network architecture and hyper-parameters have not been optimized for predicting $B_z$ in this paper so the results should be interpreted as simple baselines. 


\subsubsection{Baseline model, features, and metrics}\label{subsubsec:models}
The Gaussian Naive Bayes (GNB) classifier is used as a baseline and implemented with Scikit-learn \cite{scikit-learn} in order to compare with the deep CNN approach. The Naive Bayes classifier assumes that the value of a particular feature is independent of the value of any other feature, ignoring any possible correlations. The advantage of this method is that it only requires a small number of training data to estimate the parameters necessary for classification. 
In our approach, we examine combinations of five input features for each of the five experiments. 
The first four features that are selected to characterize the image pixel distribution are the max, min, mean, and standard deviation.
The last feature leverages the mathematics of shape, as applied to space weather prediction in a related but different topological approach\cite{fractal20}, to compute the Minkowski–Bouligand (MB) fractal dimension or box-counting dimension. The MB fractal dimension 
takes on values between 1 and 2 for a 2D image. 
The box-counting dimension is computed on sub-sampled SoHO images which are each first transformed into a binary representation using the respective means of these images as the discriminator. 
At each of the experiments' percentile thresholds, a best-fit model is determined. For the single SoHO product experiment, there are 31 distinct feature combinations. For the other two multi-product experiments, one feature per product is considered which yields 216 combinations for three products and 279,936 combinations for all seven products, of which one in ten is sampled.
The GNB approach does not utilize a validation set. As such, in order to maintain a direct comparison with the CNN results, the validation set is excluded rather than used to make a larger training set.

The predictions from the CNN and GNB are compared using the TSS and MCC scores. 
TSS is the difference between the true and false positive rates and is an unbiased estimator with respect to class imbalance\cite{camporealeML}. 
MCC is a correlation coefficient between the observed and predicted binary classifications and can be used if the classes are of very different sizes. Both measures return a value between -1 and +1. 
Each of the GNB model TSS and MCC scores can be directly compared with the same class of CNN models since the number of positives and negatives is fixed for the same products used at a given threshold.  
Expressions for TSS and MCC are given in eqs.~\ref{eq:TSS} and~\ref{eq:MCC} as

\begin{align} 
\text{TSS} &= \text{TPR} - \text{FPR} = \frac{\text{TP}}{\text{TP} + \text{FN}} - \frac{\text{FP}}{\text{FP} + \text{TN}}, \label{eq:TSS} \\
\text{MCC} &= \frac{\text{TP} \times \text{TN} - \text{FP} \times \text{FN}}{\sqrt{(\text{TP}+\text{FP})\times(\text{TP}+\text{FN})\times(\text{TN}+\text{FP})\times(\text{TN}+\text{FN})}},  \label{eq:MCC}
\end{align}
where TP, TN, FP, FN denote true-positive/negative and false-positive/negative.

\subsubsection{Results and discussion}\label{subsec:case}
In Table~\ref{tab:results_GNB}, we present baseline results arising from the application of a Gaussian Naive Bayes (GNB) classifier and a deep convolutional neural network (CNN) to a binary classification of $B_{z}$, which serves as a proxy for the onset of geomagnetic disturbances at Earth.  
The features for the GNB are a set of standard statistical quantities characterizing the image pixel value distribution (min, max, mean, standard deviation) with the addition of the Minkowski–Bouligand (MB) fractal dimension\cite{MB}. 
Additionally, there is a `dummy' feature for each product that is set to zero. The selection of this dummy feature in Table~\ref{tab:results_GNB}, indicated by a (0), shows that the respective product was not informative.

The best performing CNN model has a True Skill Statistic (TSS) of 0.30 and and Matthews correlation coefficient (MCC) of 0.24 and is obtained for the experiment with all seven SoHO products, with the F-corona left in for the LASCO products, evaluated
at the 25$^{th}$ percentile of $B_z$ corresponding to $-6.28$~nT as shown
in Table~\ref{tab:results_GNB}. 
It is interesting that with the F-corona subtracted, the model performance drops to a TSS of 0.18 and MCC of 0.20. The difference here is that there are $\sim 700$ fewer synced times available when the F-corona is subtracted as seen from Table~\ref{tab:sixhrcad}, which results in 21 fewer true events as only 65 true events (TP + FN) remain following F-corona subtraction as opposed to the 86 true events starting out when the F-corona is left in. This is a $\sim 25$\% reduction in the number of true events which may contribute to the drop in performance. 
For this same experiment, the GNB model has a TSS of 0.39 and MCC of 0.35 with the features: standard deviation of MDI, no EIT~195, 
maximum of EIT~171, standard deviation of EIT~284, standard devitation of EIT~304, maximum of LASCO~C2, and minimum of LASCO~C3. 
Although the GNB outperforms the CNN in this example, we note that the CNN's architecture and hyper-parameters have not been optimally selected. 
The GNB best-fit model with maximum for MDI, MB fractal dimension for EIT~195, minimum for EIT~171, mean for EIT~284, maximum for EIT~304, minimum for LASCO~C2, and mean for LASCO~C3 with the F-corona in the LASCO images subtracted at the 5\% threshold in Table~\ref{tab:results_GNB} has 
a sensitivity or true positive rate (TPR) of 89\% and the highest GNB specificity or true negative rate (TNR) of 89\% 
making this a satisfactory model for predicting ``all clear" scenarios for space weather applications.  
That this model has a TSS score of 0.78 but only an MCC score of 0.41 also re-affirms that a single metric is insufficient to fully describe a model's performance. 
Comparing this GNB best-fit model with the F-corona subtracted with the GNB best-fit model at the 5\% threshold with the F-corona left in, shows an added $\sim 15$\% in TSS and $\sim 24$\% in MCC performance compared to a TSS of 0.68 and MCC of 0.33. Subtracting the F-corona increases metric performance in the seven product experiments at the 5\% to 15\% thresholds.



It is likely that a future extensive architectural study combined with a hyper-parameter sweep will produce a neural network that can outperform the GNB baseline.  
Capturing temporal evolution of magnetograms may hold the key for further predicative power. 
That there is information content in the SoHO images which does allow for forecasting is encouraging to prompt further investigations by the community. 
Since the code is intrinsically agnostic to the ground truth labels, a variety of space weather parameters can be examined.

The machine learning-ready dataset and the software pipeline with an executable Jupyter notebook for ML experiments, 
such as illustrated by Fig.~\ref{fig:pipeline_step3},
presented in this paper pave the way toward establishing a community-wide practice that emphasizes reproducible research and
benchmark datasets in space weather. An advantage of this pipeline is that it is highly modular thanks to its wrapping around
both SunPy Fido and DRMS functionality. It enables a sophisticated user to add on other data providers and data products
that are supported by the SunPy community and JSOC. The user can opt to generate FITS files locally or to work with the
pre-generated HDF5 data cubes. Another advantage of the pipeline is that different down-sampling strategies may be better
suited to different SoHO image products as observed from the features of the best fit GNB models in Table 5. Filters for missing
data and planetary transits filter the data from these signal incursions which interfere with the system’s learning of the real data.
Furthermore, there is a significant compute performance advantage over having to download each and every FITS file locally in
order to determine its suitability versus querying the JSOC and SDAC databases on-the-fly to fetch suitable files and performing
the image integrity assessment locally. The code also outputs clickable URLs to all products that have been discarded due to
having missing data and separate identifiers of planetary transits. Furthermore, the scale of the observed region is kept as in the
original images (no re-scaling). This should be appropriate for most machine learning applications. However, there might be
some specialized applications, such as super-resolution, where the performance could be improved if the apparent size of the
Sun is kept constant. Such re-scaling should then be applied downstream of our pipeline. If it is eventually found that there is
not enough information contained in the images for forecasting space weather, this would still be an important step.

\section{Code and Data availability}\label{subsec:codeavail}
All of our Python code and Jupyter notebooks are open-source.
Further details on virtual environment setup, code usage with sample benchmarks, and accompanying Jupyter notebooks for 
i. performing the CNN and GNB experiments on one, three, and all seven SoHO image products, with and without the F-corona, as described in this paper and 
ii. illustrating filters for planet and comet transits as well as cosmic ray and blob detection, 
may be obtained via (\url{https://github.com/cshneider/soho-ml-data-ready}). 
Software versions of required libraries are provided in the `Mission\_Requirements' text document and can be automatically installed with the specified conda environment using `conda install' as in the instructions of the aforementioned GitHub repository. 
Interested parties are encouraged to get involved in the ongoing developments of the dataset.

\bibliography{refs}

\section*{Acknowledgements}
We would like to thank Fr\'{e}d\'{e}ric Auch\`{e}re of Universit\'{e} Paris-Sud 11, Institut d'Astrophysique Spatiale, Orsay, France, for his advice with EIT. 


We are grateful to all the developers of the following software and machine learning software libraries: TensorFlow \cite{tf15}, Scikit-learn \cite{scikit-learn}, NumPy \cite{numpy}, Matplotlib \cite{matplotlib}, IPython: Jupyter \cite{jupyter}.

CS was partially supported by NWO-Vidi Grant No. 639.072.716.
CS would like to thank Drs. Mandar Hemant Chandorkar and Rakesh Sarma for helpful discussions in regard to TensorFlow code implementation.

KB was supported by NASA funds to the NRL in support of the SOHO/LASCO project.

EC is partially funded by NASA under Grant No. 80NSSC20K1580 (Ensemble Learning for Accurate and Reliable Uncertainty Quantification).
This project has received funding from the European Union's Horizon 2020 research and innovation program under Grant Agreement No. 776262 (AIDA).

\section*{Author contributions statement}
Carl Shneider wrote the software pipeline, visualized the software pipeline, developed and ran the CNN networks using TensorFlow, visualized the CNN network, bench-marked all of the runs, ran the Gaussian Naive Bayes experiments with features including the fractal dimension, analyzed all the results, visualized all the results (except for Fig.~\ref{fig:intensities}), and wrote the paper.\\
Andong Hu helped to develop the CNN networks using PyTorch, visualized the CNN network, run a subset of the Gaussian Naive Bayes experiments and provided analysis on these experiments, and provided insightful comments on the paper. \\
Ajay K. Tiwari produced Fig.~\ref{fig:intensities}, helped write Section~\ref{sec:techvalid}, helped incorporate FITS metadata into the HDF5 data cubes, and provided insightful comments on the paper. \\
Monica G. Bobra provided insightful information on performing queries with Fido and DRMS.\\
Karl Battams provided insightful information on the LASCO coronagraph images and edits in Section~\ref{sec:methods}. \\ 
Jannis Teunissen advised on the visualization and data analysis, and provided insightful comments on the paper and code.\\
Enrico Camporeale advised on all aspects of the project, helped write the paper, and provided insightful comments on the code.

\section*{Competing interests}
The authors declare no competing interests.

\end{document}